\documentclass[conference]{IEEEtran}

\IEEEoverridecommandlockouts
\usepackage{cite}
\usepackage{amsmath,amssymb,amsfonts}
\usepackage{algorithmic}
\usepackage{framed}
\usepackage{graphicx}
\usepackage{textcomp}
\usepackage{xcolor}
 \usepackage[colorlinks=true,
        linkcolor=black,
        citecolor=black,
        filecolor=black,
        urlcolor=black,
        bookmarks=true,
        bookmarksopen=true,
        bookmarksopenlevel=3,
        plainpages=false,
        pdfpagelabels=true]{hyperref}
\def\BibTeX{{\rm B\kern-.05em{\sc i\kern-.025em b}\kern-.08em
    T\kern-.1667em\lower.7ex\hbox{E}\kern-.125emX}}
    
\usepackage{ifthen}

\newboolean{auxinfo}
\setboolean{auxinfo}{true}   

\newboolean{showoutline}
\setboolean{showoutline}{true}

\newboolean{showcomment}

\setboolean{showcomment}{true} 

\ifthenelse{\boolean{auxinfo}}
  {}
  {
    \setboolean{showoutline}{false}
    \setboolean{showcomment}{false}
  }

\usepackage{xcolor}
\usepackage{amssymb}

\ifthenelse{\boolean{showcomment}}
   {\newcommand{\nb}[2]{\fcolorbox{gray}{yellow}{\bfseries\sffamily\scriptsize#1}{\sf\small$\blacktriangleright${\em #2}$\blacktriangleleft$}}
   \newcommand{\working}[1]{\fcolorbox{gray}{yellow}{{\bf #1}\emph{\scriptsize---in progress---}}}
   \newcommand{\TBD}[1]{\fcolorbox{gray}{yellow}{{\bf #1}\textbf{TBD}}} 
  }
  {\newcommand{\nb}[2]{}{}
   \newcommand{\working}[1]{}
   \newcommand{\TBD}[1]{} 
  }

\usepackage{todonotes}
\ifthenelse{\boolean{showoutline}}{
	\newcommand{\outline}[3]{
		~\newline 
		\fcolorbox{red}{white}{
			\parbox{\columnwidth}{
				\ifthenelse{\equal{#1}{}}{
					\ifthenelse{\equal{#2}{}}{
						\noindent\colorbox[rgb]{0.65,0.16,0}{\textcolor[rgb]{1,1,1}{\textbf{Outline}}}
					}{
						\colorbox[rgb]{0.65,0.16,0}{\textcolor[rgb]{1,1,1}{\textbf{Outline -- Responsible: #2}}}
					}
				}{
					\ifthenelse{\equal{#2}{}}{
						\noindent\colorbox[rgb]{0.65,0.16,0}{\textcolor[rgb]{1,1,1}{\textbf{#1 page(s)}}}
					}{
						\colorbox[rgb]{0.65,0.16,0}{\textcolor[rgb]{1,1,1}{\textbf{#1 page(s) -- Responsible: #2}}}
					}
				}
				#3
			}
		}
	}
}{
	\newcommand{\outline}[3]{}
}

\newcommand\defauxcomm[1]{
       \expandafter\newcommand\csname #1\endcsname[1]{\nb{#1}{##1}}
       \expandafter\newcommand\csname WK#1\endcsname{\working{#1}}
       \expandafter\newcommand\csname TBD#1\endcsname{\nb{#1}}
    } 
   
\defauxcomm{KS}
\defauxcomm{MN}

\usepackage[normalem]{ulem}
\ifthenelse{\boolean{showcomment}}
    {\newcommand{\strike}[1]{\textcolor{red}{\sout{#1}}}}
    {\newcommand{\strike}[1]{}}

\begin{document}

\title{How to Measure Performance in Agile Software Development? A Mixed-Method Study}

\author{\IEEEauthorblockN{Kevin Phong Pham}
\IEEEauthorblockA{University of Applied Sciences FHDW Hannover \\
Hannover, Germany \\
kevin\_phong.pham@edu.fhdw.de}
\and
\IEEEauthorblockN{Michael Neumann}
\IEEEauthorblockA{University of Applied Sciences and Arts Hannover \\
\textit{Dpt. of Business Information Systems}\\
Hannover, Germany \\
michael.neumann@hs-hannover.de}
}


\maketitle

\begin{abstract}
\textit{Context:} Software process improvement (SPI) is known as a key for being successfull in software development. Measuring quality and performance is of high importance in agile software development as agile approaches focussing strongly on short-term success in dynamic markets. Even if software engineering research emphasizes the importance of performance metrics while using agile methods, the literature lacks on detail how to apply such metrics in practice and what challenges may occur while using them. \textit{Objective:} The core objective of our study is to identify challenges that arise when using agile software development performance metrics in practice and how we can improve their successful application. \textit{Method:} We decided to design a mixed-method study. First, we performed a rapid literature review to provide an up-to-date overview of used performance metrics. Second, we conducted a single case study using a focus group approach and qualitativ data collection and analysis in a real-world setting. \textit{Results:} Our results show that while widely used performance metrics such as story points and burn down charts are widely used in practice, agile software development teams face challenges due to a lack of transparency and standardization as well as insufficient accuracy. \textit{Contributions:} Based on our findings, we present a repository of widely used performance metrics for agile software development. Furthermore, we present implications for practitioners and researchers especially how to deal with challenges agile software development face while applying such metrics in practice. 
\end{abstract}


\section{Introduction}
\label{Sec1:Intro}

In the evolving field of software development, agile methods like Scrum, Kanban, and eXtreme Programming have become essential~\cite{VersionOne.2023}. 
They provide the flexibility and adaptability agile development teams need to meet changing requirements and customer demands, which are triggered by the increased dynanism of the markets~\cite{Bennett.2014}. Today, agile methods used worldwide in various different contexts and are thus often understood as state of the art approaches in software development. 

We know, that the ability of agile software development teams to react to new or changed circumstances is one of the major objetives for companies to use agile methods in practice (e.g., \cite{VersionOne.2023}). 
While using agile methods measuring teams' performance becomes particularly essential because it fosters transparency within the team~\cite{almeida2023perceived,soini2011survey}. This transparency is of high importance for the iterative process of inspection and adaptation that agile approaches emphasize and on which the process improvement relies on.
By continuously monitoring and evaluating key performance metrics, teams are able to identify areas where processes can be improved, inefficiencies can be eliminated, and overall productivity can be increased~\cite{almeida2023perceived}.
This iterative approach of inspect and adapt allows agile software development teams to foster continuous improvement (which we also know as Kaizen) leading to central objectives using agile methods in practice, e.g., stay aligned with customer needs and project objectives effectively. 
Performance metrics serve as feedback mechanisms that inform the team whether their adaptations are moving the project in the right direction~\cite{paulish1994case}.

We know that effective measurement and analysis of these metrics are fundamental realizing the potential of agile methods in practice, ensuring that the teams adapt quickly and continuously steers the project towards its goals~\cite{soini2011survey}. Thus, agile software development necessitates the use of suitable performance metrics measure and enhance team performance and efficiency. 

Previous research~\cite{Choras.2020,Salido.2023} shows that performance metrics are crucial for understanding, predicting, and evaluating software development projects.
Existing software metrics have been widely studied, and recent studies have discussed their reasons for use and effects within ASD~\cite{kupiainen2014industrial}.
Choosing and using the right metrics can facilitate early problem detection and enhance decision-making within teams.
Conversely, using inappropriate metrics can introduce biases and lead to undesirable behaviors.~\cite{soini2011survey}).

Current knowledge and use of performance metrics in ASD face several challenges. For example, the high variety in estimation techniques~\cite{Neumann.2021} or methods provides flexibility for agile software development teams~\cite{kupiainen2015using,usman2014effort,lopez2022quality}
In turn, this situation also leads to an increased complexity ensuring the selection of the most appropriate metric for a given context. 
This can lead to inconsistencies in data quality if the chosen method does not align well with the specific project requirements or user capabilities.
The comparability of teams is restricted due to the use of different metrics, as the absence of uniform structures allows for varied approaches to measurement. 
Additionally, the variability in methods can complicate training and standardization efforts across an organization, potentially resulting in misinterpretations and misalignment in strategic objectives.

\newpage 
Thus, the above motivates the objective of our study, which is refined by the following research questions:
\begin{itemize}
    \item \textbf{RQ1:} Which metrics are used in agile software development teams to measure performance? 
    \item \textbf{RQ2:} What are real-world challenges agile software development teams face when using performance metrics?
    \item \textbf{RQ3:} What specific performance metrics can be used and optimized in practice to measure project success?
\end{itemize}

The paper at hand is structured as follows: In Section~\ref{Sec2Background}, we give a brief introduction on metrics in the area in agile software development and further provide an overview of the identified work related to our studies topic. Section~\ref{Sec3:Research Design} describes our research design. In Section~\ref{Sec4:Results}, we present the results of our study followed by a discussion to present practical implications in Section~\ref{Sec5:Implications}. Before the paper closes with a summary in Section~\ref{Sec7:Conclusion}, we outline the limitations of the study in Section~\ref{Sec6:ThreatstoValidity}.

\section{Background \& Related Work}
\label{Sec2Background}

\subsection{Metrics in Agile Software Development}

Metrics are quantitative measures used to assess, quantify, and monitor various aspects of systems, processes, products, or performances~\cite{heimann2010metrics,kaner2004software}. 
In ASD, metrics are of high importance role in measuring the performance, quality, and progress of development processes~\cite{misra2011survey}. 

However, agile approaches with their iterative-incremental characteristic differ significantly from phase-oriented process models like the Waterfall approach with the result at the end of the project. Thus, several authors describe agile methods as a reaction to established phase-oriented (or big design upfront~\cite{Sidky.2007}) approaches (e.g., \cite{Abrahamsson.2002,Williams.2010}. For the metrics used in software development, this fundamental led to the situation that established metrics used in traditional settings are not fully transferable to agile environments~\cite{kunz2008software}. 

Phase-oriented approaches often employ rigid, predefined metrics focused on process adherence and milestone achievement. 
In contrast, agile methods utilizes more flexible, iterative metrics that evolve with the project, emphasizing both product quality and process efficiency~\cite{misra2011survey}). 
Metrics in phase-oriented software development are typically quantitative and reviewed at the end of phases (or the beginning of new ones). 
In turn, agile metrics include both qualitative and quantitative data and are updated more frequently to reflect ongoing feedback and adjustments~\cite{misra2011survey}.

This discrepancy highlights the need for developing specific metrics tailored to an agile method in use, a need driven by the increasing popularity of agile approaches~\cite{Tarhan2014}).
As described in literature, metrics provide objective information that enables precise and comparable assessments~\cite{misra2011survey}.
They convert abstract concepts into measurable units, which helps in identifying successes or challenges. Metrics also allow continuous monitoring and control of processes and performance, facilitating the tracking of developments over time~\cite{misra2011survey}.
Additionally, metrics serve as a foundation for informed decision-making by highlighting problems, identifying improvement opportunities, and developing optimization strategies~\cite{misra2011survey}.
Moreover, different metrics can challenge team members to achieve better results, potentially leading to behavioral changes within the team~\cite{misra2011survey}.

Metrics can be grouped into several categories, each with a specific purpose for the tracking and analysis of certain aspects of a project. Some of the main categories are e.g. process metrics, product metrics, quality metrics or risk metrics~\cite{mills1998metrics}. 
Each of these main categories can be further distinguished into more specific sub-categories to provide more detailed insights into particular areas or processes.
In their comprehensive study, Usman et al.\ employed a taxonomy design method to systematically and precisely organize various types of metrics~\cite{usman2017effort}. 
This taxonomy supports structuring the metrics to ensure they are categorized accurately, facilitating better analysis and utilization in software development projects.
Performance metrics are quantifiable measures used to evaluate the effectiveness, efficiency, and quality of various aspects of an operation, process, or system.
These metrics are crucial for assessing performance, guiding decision-making, and identifying areas for improvement.
Previous research and reviews provide a wide variety of performance metrics available, reflecting the diverse needs and goals of different projects and industries~\cite{kupiainen2014industrial,usman2017effort}).

\subsection{Related Work}
ASD has gained significant research interest in recent years as organizations seek to enhance their adaptability and deliver high-quality software solutions in rapidly changing environments~\cite{almeida2023perceived}. 

Extensive research has examined the benefits of agile methodologies and metrics across various industries (Misra and Omorodion, 2011 \cite{misra2011survey}). 
However, there are notable gaps in the literature, especially regarding the application and effects of these methods in regulated sectors such as insurance and pensions. 
Research in these areas has been limited and focused specifically on certain corporate contexts (Choras et al., 2020 \cite{choras2020measuring}; Pichler et al., 2006 \cite{pichler2006agile}).

Systematic literature reviews on agile software development have typically concentrated on general best practices, often neglecting the unique challenges of the insurance industry (Kupiainen et al., 2015 \cite{kupiainen2015using}; Nguyen and Tran, 2013 \cite{nguyen2013review}). 
Research primarily focuses on the adoption of agile methods~\cite{Neumann_2024} and the use of metrics to measure their impact~\cite{Topp_2022}, with few studies examining how these practices are applied in real-world scenarios, particularly within sectors like insurance that require stringent data integrity and security. 
This research seeks to address these gaps by specifically analyzing the application, challenges, and benefits of performance metrics in agile software development within the insurance sector through a mid-sized german company.

\section{Research Design}
\label{Sec3:Research Design}
We selected a mixed-method research design to gain an in-depth understanding of metrics in agile software development. Our research design consists mainly on two different approaches a) a rapid literature review and b) a single case study using qualitative data collection and analysis methods. Both research methods were designed, prepared, and conducted using guidelines. In this section, we first explain our research design, followed by a detailed explanation of the applied research methods in the subsections below. 

\begin{figure*}[htbp]
\centering
\includegraphics[scale=0.3]{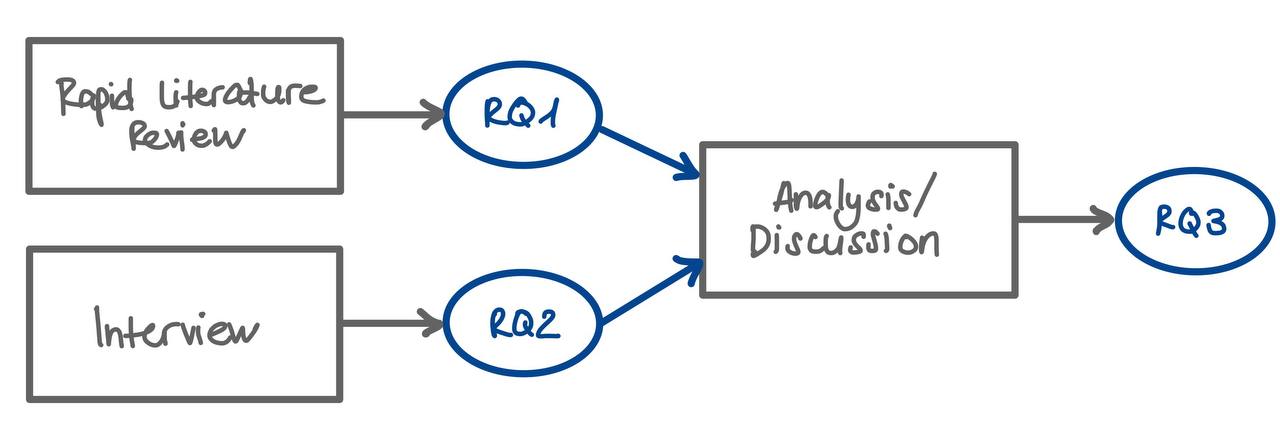}
\caption{Mixed-method research approach}
\label{fig1}
\end{figure*}

Figure~\ref{fig1} depicts our research design and the connection answering our three research questions. Initially, a rapid literature review was conducted to identify, describe and evaluate documented agile metrics answering RQ1. 
This phase is essential for the collection of best practices and scientific findings.
To investigate the second and third research question, an assessment of the current state was then conducted two different focus groups in a mid-sized company operating in the insurance sector. 
These interviews are used to determine satisfaction with current working methods and processes and to gain industry-specific insights. 
Recommendations for potential optimizations are finally developed based on these findings, ultimately leading to suggestions for improvements that address subsequent research questions. 

\subsection{Rapid Literature Review}

Rapid Literature Reviews (RLR) are streamlined versions of systematic literature reviews, designed to quickly synthesize available research within a condensed timeframe~\cite{cartaxo2020rapid}.
By focusing on studies and summarizing evidence more succinctly and efficiently, RLRs provide timely insights into specific topics of interest often with a dedicated context defined e.g., based on research questions. This approach can efficiently capture and apply current knowledge to improve processes and performance metrics. 

We designed, prepared, and conducted the RLR mainly based on the guideliens by Cartaxo et al.~\cite{cartaxo2020rapid}. The RLR method followed a three step approach: a) Defining a search strategy including search strings and select scientific search engines, b) defining a study selection strategy applying inclusion and exclusion criteria, and c) data extraction and analysis of the final result set. However, we took some adaptions of the guidelines~\cite{cartaxo2020rapid}. We give more information below, when we explain the review process on detail. 

Key terms were strategically selected based on the research questions to ensure coverage of relevant aspects.
The terms chosen included ”agile”, ”software”, ”metric”, ”performance”, ”mapping”, ”systematic”, and ”taxonomy”.
To optimize the result set by applying a correct search string, we performed test runs using the search engine from ACM Digital Library to refine the search string. In total, we created three final search strings: 
\begin{itemize}
    \item ”agile AND metric AND performance AND taxonomy”
    \item ”software AND taxonomy AND mapping”
    \item ”agile AND metric AND systematic”
\end{itemize}

To capture a broad spectrum of relevant primary and secondary studies, we conducted our search in two scientific databases: ACM Digital Library and ScienceDirect. The final search runs were conducted on 20.11.2023 resulting in a total of 43,088 studies. The result set was narrowed down by using filter settings to apply the inclusion and exclusion criteria related to the systematic characteristics of the studies such as the article type (research articles) and the year range filter (max. 10 years old studies; since 2013) to capture recent trends in the field. After applying the filter settings the final result set consisted of 24,126 studies. 

First and foremost, it is worth to mention that we had to handle a massive set of potential relevant studies. As this study is designed as mixed-method study and the RLR was selected because of its characteristic to summarize evident knowledge in an efficient way, we decided to focus on the most relevant results. Thus, we focused the literature review including the study selection process on the first 25 papers per search run (150 studies in total). However, it is worth to mention that even we screened the other title of the rest of the studies from the result set, we did not find other relevant studies for our RLR. For the result set of the 150 included studies, we performed a three-step selection process—starting with title evaluation, followed by abstract analysis, and culminating in full-text review based on relevance and depth of insight—only 5 scientific papers were ultimately selected for detailed analysis in this study. Further information related to the RLR including the protocoll of the study and the selection process are made available at Zenodo~\cite{Pham.2024-2}.

\subsection{Single Case Study}
We decided to conduct a single case study using a qualitative research approach as we want to create an in-depth understanding of the application of performance metrics in agile software development. For the design and preparation of the study, we used the guidelines from Runeson and Hoest~\cite{Runeson.2009} and Yin~\cite{Yin.2009}.  As shown in Section~\ref{Sec2Background}, research results emphasize the importance and benefits of using metrics in agile software development covering different industries (e.g., presented by Misra and Omorodion~\cite{misra2011survey}). However, gaps remain in the literature, particularly concerning their application and impacts in regulated sectors such as insurance and pensions.
Consequently, this study focused on examining the use and effects of these metrics within a German mid-sized software development company operating in the insurance sector. Here, we use the anonymized name \textit{Dunder Mifflin Inc.} for the case company. Further information about the case company can be provided upon request and after a proof by the case company. 
Dunder Mifflin Inc. provides a unique opportunity to understand how agile practices and metrics perform in an environment characterized by stringent regulatory requirements and complex risk management needs. 

As mentioned above, the second research method covered qualitative data collection and analysis using two focus groups applying semi-structured interview sessions. Here, we aimed gathering the data with key focus groups within agile software development teams in the case company, selected based on their central roles using performance metrics. 
The first focus group included Software Developers of various experience levels, providing diverse perspectives on metric application. In the second focus group we interviewed Product Owners, who offer insights into how metrics influence decision-making and product planning. An overview of the participants profiles is given in Table~\ref{tab:charakteristikaInterview}.

\begin{table}[h]
\centering
\caption{Characteristics of the participants per focus group}
\label{tab:charakteristikaInterview}

\begin{tabular}{|l|c|c|}
\hline
 & \textbf{Developer} & \textbf{Product Owner} \\ \hline
Sample size (n) & 6 & 2 \\ \hline
Average experience in years & 11.75 & 12.5 \\
Average employment period & & \\ 
in years at the company & approx. 8.5 & 12.5 \\ \hline
Average agile experience in years & approx. 4.5  & 5  \\  \hline
\end{tabular}
\end{table}

This approach allows an in-depth understanding of group dynamics and individual experiences related to performance metrics in agile settings. 
The interview questions were meticulously designed to allow participants from the case company to openly share their experiences, focusing on issues and challenges related to the use of performance metrics. 
The questions covered various aspects of using performance metrics, including type, frequency and relevance of metrics used as well as existing challenges and success stories.
Attention was paid to asking open-ended questions that provided scope for detailed responses and captured the various perspectives of the participants. The interview guideline used is available at Zenodo~\cite{Pham.2024-1}.

We refrained from recording the focus group dates for three reasons: a) Regulatory circumstances in the case company and the resulting consequences for later publication, e.g. through blocking notices, b) Potential effects of the participants with regard to their openness and transparency during the interviews and c) Recordings of group discussions are difficult to transcribe because, for example, it is not always clear who is speaking. Both focus group appointments were conducted in German and onsite in the office of the case company. This data collection approach made informal talks possible and we further were able to consider non-verbal communication. Both focus group were held in January 2024 and took between 30 to 45 minutes, moderated by the first author.  

We performed the data analysis using the field notes and manually created protocols of the focus group sessions. All the data was documented using Microsoft Excel sheets and notes in a textfile. In the first analysis step, we merged the data to one Microsoft Excel sheet and categorized the data based on the interview guideline. The data analysis was performed using open coding aiming to reduce the complexity of the data. The open coding process led to a comprehensive, but more systematic data basis which we then analyzed in an coding to identify categories. 

This two-step data analysis approach enabled a comprehensive analysis of the current state of metrics used in the teams in which the focus group participants operate.
Considering ethical reflections related to our qualitative research approach including humans, their behaviour and beliefs, we took several actions. First, we guarantee the anonymity of each participant of the study and the case company including all relevant information such as organizational units, software development projects, or even products.   Pre-arrangements were made with volunteer participants to ensure structured interviews and we took the consent by all the participants to be allowed taking field notes of the interviews. Thus, participants responses were briefly transcribed by taking notes during the focus group sessions. 

\section{Results}
\label{Sec4:Results}

\subsection{Repository of Performance Metrics in Agile Software Development}
In this subsection, we present the findings from our rapid literature review and answer our first research question: \textit{Which metrics are used in agile software development teams to measure performance?}

\begin{table}
\centering
 \caption{Overview of the included studies}
  \label{tab1:OverviewofRelWork}
  \begin{tabular}{|l|l|c|c|}
\hline
Publication [Ref.] & Research design & Year & Database \\
\hline
S1~\cite{usman2014effort} & Systematic literature review & 2014 & ACM\\
\hline
S2~\cite{chloros2022use}  & Systematic Mapping Study & 2022 & ACM\\
\hline
S3~\cite{ram2019success}  & Multiple Case Study & 2019 & ACM\\
\hline
S4~\cite{kupiainen2014industrial}  & Systematic literature review & 2015 & SD\\
\hline
S5~\cite{lopez2022quality}  & Systematic Mapping Study & 2022 & SD\\
\hline
\end{tabular}
\end{table}

Following the previously described selection process of the current literature, a total of five studies were retained for detailed analysis. 
Each of the identified studies (Table~\ref{tab1:OverviewofRelWork} provides an overview) provides a comprehensive overview of methods and metrics used in agile software development. 
However, there are remarkable differences between the identified metrics in several aspects. 
A thorough analysis of the metrics listed in these studies led to the development of a repository of performance metrics, which documents the metrics in detail and is available at Zenodo ~\cite{Pham.2024}. 
It provides a clear understanding and comparison of agile software development metrics highlighting similarities and differences. 
Commonly referenced metrics included Story Points, Function Points, and Velocity, illustrating their prevalence in agile methods~\cite{kupiainen2015using,lopez2022quality,usman2014effort}. 
The variety of metrics presented demonstrates the adaptability of agile methods to various requirements in different contexts. Covering established metrics like Story Points or Function Points, but also more specific agile metrics such as Custom Metrics and Lead Time, there is a broad spectrum available for performance evaluation. 
The studies collectively offer an extensive collection of metrics customized specifically for measuring work effort, showcasing the flexibility and depth of metric usage in agile environments.

Based on our analysis, we can highlight Velocity as a key metric for estimating workload in planning future iterations, emphasizing its role in measuring the efficiency and performance of agile software development teams~\cite{kupiainen2015using,ram2018software}). 
Its widespread use and examination in these studies underline Velocity's important function in planning and managing agile development processes.

Furthermore, metrics specifically designed to assess the accuracy of estimations were prominently featured. 
The study by~\cite{usman2014effort} emphasize the relevance these metrics serve in evaluating how close initial estimates are to the current outcome or output. 
This capability is instrumental in refining the estimation processes used in iteration planning meetings.

The analysis also highlights that several metrics pertain specifically to the process of error identification and correction~\cite{lopez2022quality}. 
This focus reflects the complexity and multifaceted nature of evaluating the performance of agile software development teams. 
The variety of metrics related to bug fixing emphasizes the importance of continuous improvement and adaptability to unforeseen challenges in agile methods.

\begin{framed}
\noindent\textbf{Answering RQ1:} Based on the identified studies we created a comprehensive repository of performance metrics used in agile software development, showcasing a wide range of metrics such as Story Points, Function Points, Velocity, Custom Metrics, Test Coverage, and Lead Time. 
\end{framed}

\subsection{Challenges while Using Performance Metrics in Practice}
Based on the results from our focus group study, we answer the second research question; RQ2: \textit{What are real-world challenges agile software development teams face when using performance metrics?}

The interview involved developers from two different teams at the company, so overlapping responses were observed, especially within the same team where similar metrics are used, and perceptions of problems and challenges are alike. These interviews provide valuable insights into the practical application of metrics from the developers' perspective, forming a crucial foundation for further discussion and analysis for both RQ2 and RQ3.

The findings reveal that developers at the company utilize a range of metrics and tools in descending order of frequency:

\begin{itemize}
    \item Story Points
    \item Four-Eyes Principle
    \item Build Status
    \item Test Coverage
    \item Person-Days
\end{itemize}

Additional metrics such as Runtime Stability, Test Success Rate, Code Quality, Code Smells, and MQ Service Status are also applied. 
Metrics are typically measured initially in the development process. For instance, Story Points are estimated during the Backlog Refinement every two weeks, where all cards in the backlog are discussed and estimated through oral Planning Poker.
Estimates are made as soon as a requirement arises and metrics are adjusted only in exceptional cases throughout the development cycle. 
The teams primarily use tools like Grafana Dashboard, SonarQube, Jira, and Confluence to measure metrics and monitor progress. 
This setup underscores the structured yet flexible approach to metric usage in agile environments at the company, highlighting both commonalities and variations in experiences and methodologies among the developers.

However, several issues and challenges related to the application of metrics and estimation methods were identified. 
A recurring issue was the perceived irrelevance or inadequacy of Story Points in some teams, which was attributed by some developers to the lack of commitment to sprints and the absence of Velocity measurement. 
The oral Planning Poker method was frequently mentioned as insufficiently detailed, and there was confusion about the exact process and estimation method. 
Developers also reported difficulties using Burn-Down Charts and estimating Story Points for cards. 
Additionally, the order situation was described as unpredictable, leading to disrupted workflows and changes in prioritization, necessitating adjustments to estimates during the development cycle. 
This resulted in discrepancies between the initial estimates and the actual time and effort required. 
It was emphasized that the accuracy of estimates should be reviewed at the end of each sprint to address these discrepancies effectively. 

In addition the interviews with Product Owners (POs) at the company provided a broader perspective on the organization's agile methods and practices. 
POs core competency rely in defining requirements and prioritizing them, offering valuable insights into the product management wrt.\ to the agile software development within the organization. Thus, they enabling a more comprehensive assessment of the software processes in use. 
Technical requirements are developed through close interactions between various teams and committees (e.g., communities of practice), with a focus on capturing and evaluating professional needs. 
Nevertheless, we identified highlighted issues with the precise definition and scope of action of POs, particularly regarding their role and clarification of responsibilities.
The evaluation of the use of Story Points presented a mixed opinion. 
Generally, the method is considered understandable and effective for estimating the scope and effort of tasks. 
However, concerns about the transparency and accuracy of these estimates were raised, particularly regarding the influence of oral estimates during Planning Poker. 
An overview of the development process is achieved through regular exchange formats like weekly meetings and the use of the Jira board, with Story Points serving as a unit of measure to track progress and capacity.
Regarding quality metrics, the focus is on domain tests, system stability, and performance. 

The respondents emphasized the importance of regular interactions with developers to ensure the quality of software solutions. 
User feedback and the continuous improvement of systems also play a crucial role. Challenges in the agile process were particularly identified in requirements management and the clarification of responsibilities. 
Issues such as lack of standardization and occasional transparency problems during ticket handover were cited as causes for conflicts and misunderstandings. 
The POs expressed a desire for clearer definition of responsibilities and improved information exchange between teams to enhance efficiency and transparency in the development process.

\begin{framed}
\noindent\textbf{Answering RQ2:} Qualitative interviews with developers and product owners at the company highlighted the structured yet flexible use of a range of metrics and tools, including Story Points, SonarQube, and Test Coverage, to monitor agile development processes. 
Key issues identified include the perceived inadequacy of Story Points and challenges with the clarity and accuracy of estimation methods like oral Planning Poker, leading to discrepancies in workload estimates and project management. 
Product owners emphasized the importance of defining roles clearly and improving communication between teams to enhance project efficiency and transparency.
\end{framed}

\section{Practical Implications}
\label{Sec5:Implications}
In this section, we first discuss practical implications based on the findings presented in the previous section aiming to answer our third research question; RQ3: \textit{What specific performance metrics can be used and optimized in practice to measure project success?}

Analyzing the results of the rapid literature review and the interviews allows to understand the utilization and optimization potential of performance metrics in agile software development teams, including those at the investigated company. 
It was found that while not all metrics described in the literature are implemented at the company, due to the vast variety available, relevant metrics are actively integrated into the software development process. 
This indicates that the selection and implementation of metrics are based on a careful balancing process, tailored to meet the specific needs and contexts of the individual development teams. 
The analysis of the results of the rapid review provides a guideline to highlight areas for improvement and suggest possible adjustments or additions to metric usage. 
This comparison between the findings from the literature and the actual practices at the company serves as a valuable foundation for identifying best practices and developing recommendations to optimize the use of metrics in agile software development.

The use of Story Points is a common practice in agile software development teams for estimating the effort required for tasks and user stories, as frequently mentioned in the literature and interviews. 
This method allows for a relative assessment of tasks, enabling teams to gauge the effort of a task in comparison to others without relying on absolute time estimates like person-days. 
Story Points enhance agility, offering adaptability to changes and uncertainties, and allowing teams to respond flexibly to new information or changes in requirements. 
However, a notable disadvantage is the lack of objectivity in Story Points, which can lead to inconsistencies and uncertainties across different teams. 
This variability challenges the comparability of Velocity between teams and could potentially hinder the scalability of agile practices at the enterprise level. 
Furthermore, if Story Points do not correlate well with Velocity or are not considered binding, it can affect the predictability of work progress. 
Both developers and product owners at the company have criticized the inadequate determination of Velocity, highlighting this as a significant issue.

Although Velocity is often identified as a key metric for measuring team performance in agile software development, it is not actively utilized within the investigated case company. 
Velocity is typically used to gauge the amount of work completed (or also described as the acceleration of a team), e.g. number of Story Points, within a specific timeframe, like an iteration.
Its measurement can provide planning certainity for future sprints, helping teams predict their capacity and set realistic goals by understanding the average amount of work completed per sprint. 
This may enable efficient resource utilization and continuous improvement as teams can adjust their used software development processes and methods based on identified trends. 
However, Velocity's effectiveness can vary significantly between teams due to different definitions of "done" and varying levels of efficiency. 
This variability can make comparing Velocity between teams challenging and may lead to manipulation of Velocity figures by teams wanting to appear more or less productive. 
Despite these potential drawbacks, Velocity can be a valuable metric if interpreted correctly and integrated into the broader context of the agile development process. 
It should not be the sole indicator of project success but rather a tool for ongoing improvement.

Planning Poker as described in the literature and also used in agile teams at the company promotes collaboration and unbiased estimates by allowing open discussion and maintaining anonymity. 
The method enhances team collaboration by facilitating open discussions on estimates, allowing for the integration of diverse perspectives, which helps in achieving more accurate workload assessments (Moløkken-Østvold et al., 2008). 
However, its oral implementation at the company could lead to uniform estimates influenced by prior responses, potentially reducing accuracy. 
The sequential approach could diminish the impact of varying perspectives and levels of experience. It also allows less room for individual contemplation and reflection, as members must immediately respond to their colleagues' estimations. 
Despite criticisms from developers and product owners regarding its execution, the fundamental practice of Planning Poker is still positively regarded within the team, suggesting that adjustments to its implementation could enhance its effectiveness.
Using Planning Poker cards could facilitate discussions and enhance all team members' involvement in estimation. 
Structured and moderated discussions ensure that all team members express their views and that diverse perspectives are adequately considered, leading to more accurate and agreed-upon estimations~\cite{molokken2008using}.

Expert Estimations is recognized as a common metric in literature and is frequently used at the investigated company, particularly in release planning. 
This method allows experienced experts to use their knowledge and skills to provide qualitative assessments of work efforts, leading to realistic and high-quality estimates, especially when the experts are familiar with the specific requirements and context. 
However, challenges such as varying opinions among different experts can lead to disagreements and uncertainties. 
Additionally, compared to quantitative methods, Expert Judgement might be less objective and transparent, potentially affecting the traceability of estimates. 
The reliance on individual expertise can also introduce biases, particularly in the absence of clear guidelines or standardized procedures. 
Furthermore, the rapid review revealed various other metrics for estimation, including advanced techniques such as neural networks.

The literature generally views metrics positively as they help measure, optimize, and manage various processes in ASD. 
Metrics are utilized for several reasons, including enhancing efficiency, tracking progress, planning future sprints, and measuring customer satisfaction~\cite{chloros2022use}. 
It is important to note that implementing improvements through metrics can entail additional work, including training, process adjustments, and monitoring new metrics. 
Thus, a careful balance between the potential benefits and the extra effort is necessary to ensure that the changes provide genuine value to the ASD teams.

\begin{framed}
\noindent\textbf{Answering RQ3:} 
The application of Planning Poker and Expert Estimation could benefit from adjustments to improve objectivity and reduce bias, such as integrating anonymized estimation processes and standardizing definitions of completion across teams.
Enhancing the use of Velocity by ensuring it is more consistently applied and reviewed can also aid in better sprint planning and resource allocation. 
Ultimately, refining these metrics and their implementation methods can lead to more accurate estimations and more efficient project management in agile environments.
\end{framed}

\section{Threats to Validity}
\label{Sec6:ThreatstoValidity}
Though we followed a systematic research design using guidelines for the rapid literature review~\cite{cartaxo2020rapid} and the focus group~\cite{Runeson.2009}, some limitations need to be taken into account. We discuss the limitations of our study using the threats to validity concept, focusing on the measures we took to address the threats to validity below.

\textit{Construct validity:} A major limitation in literature applies related to the completeness of the identified result set. However, we countered this aspect using a systematic approach with the applied guidelines for a rapid literature review. Furthermore, the study selection was performed by applying defined inclusion and exclusion criteria and a cross-check of the selection by the second author. 

Also, construct validity threats apply for the qualitative focus group approach. First, the length of the focus group sessions took between 30 and 45 minutes. Such an appointment can be tiring in the long run and thus, affect the quality of the results. For this reason, we did not hold the focus group meetings at off-peak times during the working day, but during core working hours. Sufficient breaks were also planned and implemented. 

\textit{Internal validity:} Although we prepared the focus groups based on a systematic literature search in the form of a rapid review, there are some limitations that need to be considered to strengthen the chain of evidence. We took three actions to reduce the risk of bias. First, the guiding questions for the focus group were non-leading questions so as not to induce any implicit direction (e.g., positive or negative affect of metrics application) in the participants. Second, the flow of the leading questions was semi-structured, which allowed us to go deeper into the direction the participants were seeking. Third, we considered a mix of roles, expertise, and experience in the composition of the focus group participants (see Section~\ref{Sec3:Research Design} and Table~\ref{tab:charakteristikaInterview} for further details). 

Furthermore, we did not record the focus group workshops in order to encourage participation and active involvement during implementation. We are aware that this may have a negative impact on the data quality, so we systematically documented the results and took detailed notes also for the informal talks around the workshops. The focus groups were prepared and moderated by the first author. The second author was involved in reviewing the analysis results.

\textit{External validity:} The external validity of the study could be improved by including additional focus groups from the case company or additional cases from other industries and regions. It would also be useful to take agile software development teams with different agile maturity levels under study, as we assume that effects on the selection and successful integration of metrics may apply.

\section{Conclusion \& Future Work}
\label{Sec7:Conclusion}
In this paper, we present the results of our mixed-method study dealing with performance metrics in agile software development. In particular, our study dealt with three research questions. Below, we conclude our findings based on the research questions, before we give a brief overview of our planned future work activities. First, the results from our literature review show that a wide range of metrics in agile software development is used. For example, several metrics to measure performance or software quality exist and are applied in practice. We decided to create a repository of metrics for/in agile software development to provide a systematic overview of used metrics. Second, based on the findings from our two focus groups (in a single case study), we found that some metrics (e.g., Story Points) may be inadequat for the given context or the underlying objective applying them to a team. Also, metrics for effort/complexity estimation (e.g., oral Planning Poker) are challenged by misconceptions or misunderstandings how the metric may be applied. However, we also identified that the application of metrics is of high importance as they increase transparency of the teams' outcome and progress during an iteration. Third, anlyzing the findings and results from the two other RQs, we found that it may be valueable approach to apply a metric like the Velocity measuring the progess of an agile software development team.

As our study is obviously limited by considering a single case company, we currently plan to expand the study to other highly regulated contexts like finance or public administration.


\bibliographystyle{IEEEtran}
\bibliography{references}

\begin{thebibliography}{10}
\providecommand{\url}[1]{#1}
\csname url@samestyle\endcsname
\providecommand{\newblock}{\relax}
\providecommand{\bibinfo}[2]{#2}
\providecommand{\BIBentrySTDinterwordspacing}{\spaceskip=0pt\relax}
\providecommand{\BIBentryALTinterwordstretchfactor}{4}
\providecommand{\BIBentryALTinterwordspacing}{\spaceskip=\fontdimen2\font plus
\BIBentryALTinterwordstretchfactor\fontdimen3\font minus \fontdimen4\font\relax}
\providecommand{\BIBforeignlanguage}[2]{{%
\expandafter\ifx\csname l@#1\endcsname\relax
\typeout{** WARNING: IEEEtran.bst: No hyphenation pattern has been}%
\typeout{** loaded for the language `#1'. Using the pattern for}%
\typeout{** the default language instead.}%
\else
\language=\csname l@#1\endcsname
\fi
#2}}
\providecommand{\BIBdecl}{\relax}
\BIBdecl

\bibitem{VersionOne.2023}
\BIBentryALTinterwordspacing
VersionOne and Collabnet, ``17th annual state of agile survey,'' 2023. [Online]. Available: \url{stateofagile.com}
\BIBentrySTDinterwordspacing

\bibitem{Bennett.2014}
N.~Bennett and G.~Lemoine, ``What a difference a word makes: Understanding threats to performance in a vuca world,'' \emph{Business Horizons}, vol.~57, no.~3, pp. 311--317, 2014.

\bibitem{almeida2023perceived}
F.~Almeida and P.~Carneiro, ``Perceived importance of metrics for agile scrum environments,'' \emph{Information}, vol.~14, 2023.

\bibitem{soini2011survey}
J.~Soini, ``A survey of metrics use in finnish software companies,'' in \emph{Proc. of the Intl. Symposium on Empirical Software Engineering and Measurement}.\hskip 1em plus 0.5em minus 0.4em\relax IEEE, 2011, pp. 49--57.

\bibitem{paulish1994case}
D.~Paulish and A.~Carleton, ``Case studies of software-process-improvement measurement,'' \emph{Computer}, vol.~27, pp. 50--57, 1994.

\bibitem{Choras.2020}
M.~Choraś, T.~Springer, R.~Kozik, L.~López, S.~Martínez-Fernández, P.~Ram, P.~Rodriguez, and X.~Franch, ``Measuring and improving agile processes in a small-size software development company,'' \emph{IEEE Access}, vol.~8, pp. 78\,452--78\,466, 2020.

\bibitem{Salido.2023}
M.~G. {Salido O.}, G.~Borrego, R.~R. {Palacio Cinco}, and L.-F. Rodríguez, ``Agile software engineers’ affective states, their performance and software quality: A systematic mapping review,'' \emph{Journal of Systems and Software}, vol. 204, p. 111800, 2023.

\bibitem{kupiainen2014industrial}
E.~Kupiainen, M.~Mäntylä, and J.~Itkonen, ``Why are industrial agile teams using metrics and how do they use them?'' in \emph{Proc. of the Intl. Workshop on Emerging Trends in Software Metrics}.\hskip 1em plus 0.5em minus 0.4em\relax ACM, 2014, pp. 23--29.

\bibitem{Neumann.2021}
M.~Neumann, ``Towards a taxonomy of agile methods: The tree of agile elements,'' in \emph{Proc. of the Intl. Conf. in Software Engineering Research and Innovation}, 2021, pp. 79--87.

\bibitem{kupiainen2015using}
E.~Kupiainen, M.~Mäntylä, and J.~Itkonen, ``Using metrics in agile and lean software development – a systematic literature review of industrial studies,'' \emph{Information and Software Technology}, vol.~62, pp. 143--163, 2015.

\bibitem{usman2014effort}
M.~Usman, E.~Mendes, F.~Weidt, and R.~Britto, ``Effort estimation in agile software development: A systematic literature review,'' in \emph{Proc. of the Intl. Conf. on Predictive Models in Software Engineering}.\hskip 1em plus 0.5em minus 0.4em\relax ACM, 2014, pp. 82--91.

\bibitem{lopez2022quality}
L.~López, X.~Burgués, S.~Martínez-Fernández, A.~Vollmer, W.~Behutiye, P.~Karhapää, X.~Franch, P.~Rodríguez, and M.~Oivo, ``Quality measurement in agile and rapid software development: A systematic mapping,'' \emph{Journal of Systems and Software}, vol. 186, p. 111187, 2022.

\bibitem{heimann2010metrics}
D.~Heimann, ``Metrics and databases for agile software development projects,'' IEEE Boston Reliability Society, 2010.

\bibitem{kaner2004software}
C.~Kaner and W.~Bond, ``Software engineering metrics: What do they measure and how do we know,'' in \emph{Proc. of the Intl. Software Metrics Symposium}, 2004.

\bibitem{misra2011survey}
S.~Misra and M.~Omorodion, ``Survey on agile metrics and their inter-relationship with other traditional development metrics,'' \emph{ACM SIGSOFT Software Engineering Notes}, vol.~36, pp. 1--3, 2011.

\bibitem{Sidky.2007}
A.~Sidky, J.~Arthur, and S.~Bohner, ``A disciplined approach to adopting agile practices: the agile adoption framework,'' \emph{Innovation in Systems and Software Engineering}, vol.~3, no.~3, pp. 203–--216, 2007.

\bibitem{Abrahamsson.2002}
P.~Abrahamsson, O.~Salo, J.~Ronkainen, and J.~Warsta, ``Agile software development methods: Review and analysis,'' no. 478, pp. 7--94, 2002.

\bibitem{Williams.2010}
L.~Williams, ``Agile software development methodologies and practices,'' in \emph{Advances in computers}.\hskip 1em plus 0.5em minus 0.4em\relax {Academic Press}, 2010, vol.~80, pp. 1--44.

\bibitem{kunz2008software}
M.~Kunz, R.~Dumke, and N.~Zenker, ``Software metrics for agile software development,'' in \emph{Proc. of the Australian Conf. on Software Engineering}, 2008, pp. 673--678.

\bibitem{Tarhan2014}
A.~Tarhan and S.~Yilmaz, ``Systematic analyses and comparison of development performance and product quality of incremental process and agile process,'' \emph{Information and Software Technology}, vol.~56, pp. 477--494, 2014.

\bibitem{mills1998metrics}
E.~Mills, ``Metrics in the software engineering curriculum,'' \emph{Annals of Software Engineering}, vol.~6, pp. 181--200, 1998.

\bibitem{usman2017effort}
M.~Usman, J.~Börstler, and K.~Petersen, ``An effort estimation taxonomy for agile software development,'' \emph{International Journal of Software Engineering and Knowledge Engineering}, vol.~27, pp. 641--674, 2017.

\bibitem{choras2020measuring}
M.~Choras, T.~Springer, R.~Kozik, L.~Lopez, S.~Martinez-Fernandez, P.~Ram, P.~Rodriguez, and X.~Franch, ``Measuring and improving agile processes in a small-size software development company,'' \emph{IEEE Access}, vol.~8, pp. 78\,452--78\,466, 2020.

\bibitem{pichler2006agile}
M.~Pichler, H.~Rumetshofer, and W.~Wahler, ``Agile requirements engineering for a social insurance for occupational risks organization: A case study,'' pp. 246--251, 2006.

\bibitem{nguyen2013review}
D.~Nguyen and D.~Tran, ``A review of effort estimation studies in agile, iterative and incremental software development,'' in \emph{Proc. of the Intl. Conf. on Research Challenges in Information Science}.\hskip 1em plus 0.5em minus 0.4em\relax IEEE, 2013.

\bibitem{Neumann_2024}
M.~Neumann, T.~Kuchel, P.~Diebold, and E.-M. Schön, ``Agile culture clash: Unveiling challenges in cultivating an agile mindset in organizations,'' \emph{Computer Science and Information Systems}, 2024.

\bibitem{Topp_2022}
J.~Topp, J.-H. Hille, M.~Neumann, and D.~Mötefindt, ``How a 4-day work week and remote work affect agile software development teams,'' in \emph{Proc. of the Intl. Conf. on Lean and Agile Software Development}, vol. 438, 2022, pp. 61--77.

\bibitem{cartaxo2020rapid}
B.~Cartaxo, G.~Pinto, and S.~Soares, ``Rapid reviews in software engineering,'' in \emph{Contemporary Empirical Methods in Software Engineering}.\hskip 1em plus 0.5em minus 0.4em\relax Springer International Publishing, 2020, pp. 357--384.

\bibitem{Pham.2024-2}
\BIBentryALTinterwordspacing
K.~P. Pham, ``Rapid review protocol \& documentation,'' 2024. [Online]. Available: \url{https://doi.org/10.5281/zenodo.12637754}
\BIBentrySTDinterwordspacing

\bibitem{Runeson.2009}
P.~Runeson and M.~H{\"o}st, ``Guidelines for conducting and reporting case study research in software engineering,'' \emph{Empirical Software Engineering}, vol.~14, no.~2, pp. 131--164, 2009.

\bibitem{Yin.2009}
R.~K. Yin, \emph{Case study research: Design and methods}, 4th~ed., ser. Applied social research methods series.\hskip 1em plus 0.5em minus 0.4em\relax Sage, 2009, vol.~5.

\bibitem{Pham.2024-1}
\BIBentryALTinterwordspacing
K.~P. Pham, ``Interview guideline for the focus groups,'' 2024. [Online]. Available: \url{https://doi.org/10.5281/zenodo.11206152}
\BIBentrySTDinterwordspacing

\bibitem{chloros2022use}
D.~Chloros, V.~Gerogiannis, and G.~Kakarontzas, ``Use of software and project management metrics in agile software development methodologies: A systematic mapping study,'' in \emph{Proc. of the European Symposium on Software Engineering}.\hskip 1em plus 0.5em minus 0.4em\relax ACM, 2022, pp. 25--32.

\bibitem{ram2019success}
P.~Ram, P.~Rodriguez, M.~Oivo, and S.~Martinez-Fernandez, ``Success factors for effective process metrics operationalization in agile software development: A multiple case study,'' in \emph{Proc. of the Intl. Conf. on Software and System Processes}.\hskip 1em plus 0.5em minus 0.4em\relax IEEE, 2019, pp. 14--23.

\bibitem{Pham.2024}
\BIBentryALTinterwordspacing
K.~P. Pham, ``Performance metrics repo documentation,'' 2024. [Online]. Available: \url{https://doi.org/10.5281/zenodo.11206133}
\BIBentrySTDinterwordspacing

\bibitem{ram2018software}
P.~Ram, P.~Rodriguez, and M.~Oivo, ``Software process measurement and related challenges in agile software development: A multiple case study,'' in \emph{Proc. of the Intl. Conference on Product-Focused Software Process Improvement}.\hskip 1em plus 0.5em minus 0.4em\relax Springer International Publishing, 2018, pp. 272--287.

\bibitem{molokken2008using}
K.~Moløkken-Østvold, N.~Haugen, and H.~Benestad, ``Using planning poker for combining expert estimates in software projects,'' \emph{Journal of Systems and Software}, vol.~81, pp. 2106--2117, 2008.

\end{thebibliography}

\end{document}